# Unusual magnetoresistance oscillations in preferentially oriented *p*-type polycrystalline ZrTe$_5$

M. K. Hooda and C. S. Yadav

School of Basic Sciences, Indian Institute of Technology Mandi, Mandi-175005 (H.P.) India

**ABSTRACT:**

Recently H. Wang *et al.* (arxiv:1704.00995) have reported quantum oscillation in magnetoresistance with the periodicity in logarithmic of magnetic field (B) for the p-type ZrTe$_5$. They have ascribed this type of behavior to the discrete scale invariance, resulting from Effimov bound states. We have prepared high quality stoichiometric (p-type) ZrTe$_5$ polycrystals and observed magnetoresistance (MR) oscillations, which are periodic in B. These oscillations are in contrast to usual SdH oscillations or log B dependent oscillations as observed for tellurium deficient and stoichiometric ZrTe$_5$ respectively. The MR follows the three dimensional Weyl semimetal like behavior, and Kohler's rule is obeyed at low temperatures. We obtained small cyclotron effective mass ($m^* \sim 0.05 m_e$), very high mobility of $\sim 2.2 \times 10^4$ cm$^2$/V-s and the signature of topological protected surface states in the compound. The magnetic data shows zero cusp paramagnetic susceptibility which supports the existence of topological surface states in ZrTe$_5$.

## INTRODUCTION

The ZrTe$_5$ is an important member of the group IV transition metal penta-chalcogenides. The initial studies on this compound were focused to understand the origin of peak shape anomaly in resistivity in the temperature (T) range of 60 - 145 K, large thermoelectric power (S) and positive magnetoresistance (MR) [1-18]. Recently it has been established that anomaly in resistivity arises due to bipolarity and the conduction anisotropies for electrons and holes [13, 17, 18]. The ZrTe$_5$ is semiconducting (*p*-type) and shows positive S in the perfect stoichiometric form, but tellurium off-stoichiometry and presence of defects gives rise to resistivity anomaly with metallic behavior and reversal of *S* sign at low temperatures [13,17,18]. The ZrTe$_5$ exhibits many exotic and interesting features of topological protected states. G. Manzoni *et al.* showed 3-D strong topological insulator (TI) behavior [4]. The band topology is sensitive to lattice parameters and interlayer distances, and a mere 3-4% change in interlayer distance can lead to strong TI to weak TI behavior [14]. However, G. Zheng *et al.* showed the direct evidences for 3-D Dirac semimetal phase of ZrTe$_5$ through angle dependent magnetoresistance (MR) measurements [15]. Interestingly, 3-D Weyl semimetal behavior shifts to 2-D Dirac semimetal upon application of 8 Tesla magnetic field along b-direction [5]. The compound shows topological edge states (TES) in large energy gap of 100 meV at the Brillouin zone center [14] which is also supported by the scanning tunneling microscopy (STM) experiment with bulk band gap of 80 meV at step edge [6]. The mass acquisition of massless Dirac Fermions (electrons), chiral magnetic effect, van Hove singularity near Fermi level, fractional Landau levels have also been reported in ZrTe$_5$ [7-9,17].

The most of studies undertaken regarding the topological features in ZrTe$_5$ have been performed on the bipolar tellurium (Te) deficient compounds where electrons (/holes) are dominant carriers below (/above) the resistivity anomaly. However the studies on stoichiometric *p*-type ZrTe$_5$ are limited in the number in the literature [2, 12, 18]. A. Pariari *et al.* performed magnetization and magneto-transport measurements on *p*-type ZrTe$_5$ single crystal and showed the coexistence of topological Dirac Fermions on the surface and 3 D Dirac cone state in the bulk [12]. Recently H. Wang *et al.* have reported log-periodic oscillations in the MR of stoichiometric *p*-type ZrTe$_5$ with low carrier density ($10^{15}$–$10^{18}$ cm$^{-3}$) and have shown the first time observance of the discrete scale invariance (DSI) resulting from the Effimov bound states physics in the 3-D electronic Fermionic system [18]. In the 3-D TIs, MR studies are of vital importance as MR is directly related to the Fermi surface and electronic properties of material. But the observation and detection of topological surface states (TSS) has always been a key challenge as bulk carriers also contribute to the surface transport. In case of small carrier mobility, separation of TSS from the bulk and impurities/defect states becomes very difficult. Generally single crystals are required to study the electronic transport of surface states. However, here we report unusual oscillations in the MR in the good quality preferentially oriented *p*-type polycrystalline ZrTe$_5$. Our study shows the evidence of 3 D Weyl semimetal like behavior in MR. The MR oscillations show a single predominant oscillation frequency in fast Fourier transform spectra along with the signature of TSS in the compound. Our magnetic data shows zero cusp paramagnetic susceptibility and supports the existence of TSS in ZrTe$_5$.

**SAMPLE PREPARATION AND MEASUREMNTS**

The polycrystalline ZrTe$_5$ was prepared by solid state reaction method route, by heating of the stoichiometric amount of pure (> 99.95%) elemental Zr and Te inside the evacuated quartz tube at *500 $^0$C* for *7* days and then it was slowly cooled to *200$^0$C*. The obtained material was further grounded, pelletized and sintered at *500 $^0$C* for *24* hours. It is worth to mention here that 5 % extra Te was taken to avoid/minimize Te deficiency in the compound. The X-ray diffraction (figure 1) and EDX analysis on the powdered compounds confirmed the single phase of the compound [2]. The Rietveld refinement of the XRD data using *Cmcm* space group, gave the crystal lattice parameters as *a = 3.96 Å*, *b = 14.62 Å* and *c = 13.91 Å* that are slightly higher than the reported in literature [12,13]. The X-ray diffraction data on the measured polycrystalline pellets showed the reflections from the (023), (115), (170), (191), (119) and (332) planes only, which indicate the preferential alignment of the crystal planes in the pellet.

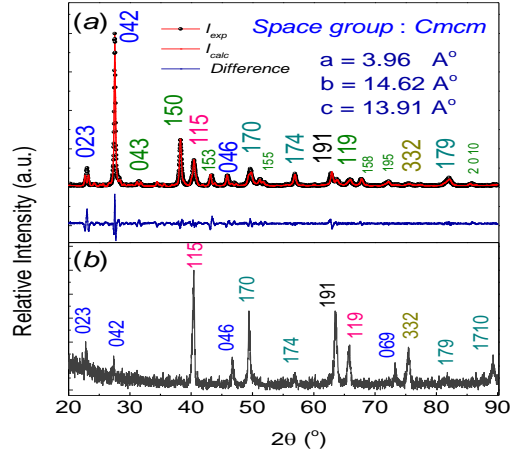

Figure 1. (**a**) Rietveld refined powder X-ray diffraction pattern of ZrTe$_5$ at room temperature (RT) (**b**) X-ray diffraction pattern of polycrystalline pellet at RT showing dominance presence of certain hkl planes.

The electronic transport measurements were performed in temperature ($T$) range of *1.8-300 K* under the magnetic field (B) of *0 - 14 T* using Quantum Design Physical Properties Measurement System (PPMS) and Quantum Design Magnetic Properties Measurement System (MPMS) was used for the magnetic measurements.

**MAGNETO TRANSPORT MEASUREMENT**

Figure 2a shows the $T$ dependence of the electrical resistivity ($\rho(T)$) measured at *B = 0, 5, 14 Tesla*. The $\rho(T)$ in our ZrTe$_5$ is similar to that reported by A. Pariari *et al.*, and does not show any peak shape anomaly down to 1.8 K. Additionally we observed positive S in the T range 1.8- 340 K [2] and positive value for the Hall carriers (see supplementary information figure S1), which indicate the high quality of our polycrystalline compounds with the dominance of one type of charge carriers (holes) only. Inset of figure 1, shows crossover from positive to negative $T$ coefficient of $\rho(T)$ at 125 K for *B = 0 Tesla* depicting metallic to semiconducting like crossover in the compound. This crossover $T$ enhances to 185 K and 250 K for *B = 5* and *14 T* respectively. Such behavior has been reported for both bipolar and *p*-type ZrTe$_5$. A slight increase in $\rho(T)$ around *T ~ 210 - 250 K*, is similar to that observed for bipolar ZrTe$_5$, and has been attributed to the finite size Fermi surface effects [10]. The MR measured at different $T$ (from *1.8 K* to *200 K*) from *-14* to *14 Tesla* (Figure 2b) is positive and shows the signature of oscillations below *15 K*. Inset of figure 2b shows the MR curves for *T = 1.8 , 3, 5, 7, 10 K*, where the oscillations are clearly visible. The undulating nature of the MR curves exhibits the signature of multiple band transport. A dip in MR near *B = 0 Tesla* and $\sim B^2$ field dependence up to *2 Tesla* at *1.8 K* is an indication of weak antilocalization (WAL) effect, which is clearly evident in the semiconducting region. The quadratic field dependence of MR is observed up to *2 Tesla* whereas linear MR dominates at higher magnetic fields. For *B > 2 Tesla*, we have fitted MR curve with the equation $y = y_0 + aB^m$ (shown in the supplementary figure S2). The value of the exponent '*m'* was found to vary from *1.1 at 20 K* to *1.45 at 200 K*. It is to mention here that MR of well-known Weyl semimetals NbP and TaP also follow similar power law dependence at high fields, with the quadratic field dependence at low fields [19, 20]. The transformation from metal to semiconducting behavior in our $\rho(T)$ data is also consistent with that predicted for 3 D Weyl semimetal [21]. The Kohler plots (MR vs *B/R*) for different *T*, shown in figure 2c, do not fall on each other for *T > 10 K*, which suggests the existence of more than one relaxation time (or scattering rates) associated with the multi-gap nature of ZrTe$_5$ [22, 23]. The merging of Kohler plots below *10 K* (inset of figure 2c) onto a single line up to *B = 3 Tesla* indicates that there is no appreciable change in carrier concentration and mobility of the compound in this field and *T* regime, which is consistent with the Hall measurement data (given in supplementary information). For *B ≤2 Tesla*, MR satisfy ~ $B^2$ criterion which is indicative of the predominantly single-band transport behavior in this compound (shown in the figure 2d). Although marked deviation from the Kohler's rule at high *T* suggests the multiband features in the compound, the transport is mainly dominated by single band with the presence of other weakly contributing bands to transport properties. The presence of several weak frequencies with single dominating frequency in Fast Fourier Transform (FFT) analysis (shown in the next section) of the oscillations at low *T*, points out towards the single predominant band behavior. Though the MR does not show saturating behavior up to *B = 14 Tesla*, its maximum value is *~75 %* at *T = 30 K*. The unsaturated linear dependence of MR with magnetic field are observed in TIs [20,24-27], Dirac semimetals [28,29], Weyl semimetals [30], and non-degenerate semiconductors compounds [31]. The quantum effects may also lead to unsaturated linear MR due to the linear energy dispersion of the Dirac fermions at the touching point [32]. The variation of resistivity with magnetic field in the extreme quantum limit for an isotropic metal can be described by Abrikosov's isotropic model [33]. This model gives rise to linear resistivity (MR) as $\rho_{xx} = \rho_{yy} = \frac{N_i B}{\pi n_0^2 e c}$ where $\rho_{ij}$ are components of resistivity tensor, $n_0$ the electron density and $N_i$ is the concentrations of scattering centers. In the case of high fields, $\rho_{xx}$ or $\rho_{yy}$ gives $n_0 \ll \left(\frac{eB}{\hbar}\right)^{3/2}$, and $T \ll \frac{eB\hbar}{m_c^*}$. For $B = 14\ T$, above conditions give $n_0 \approx 3.10 \times 10^{18}\ cm^{-3}$ and $T \approx 354\ K$ using $m_c^* = 0.05 m_e$. We can see from the above values that the conditions for quantum linear MR is satisfied even at *354 K*. J. Feng *et al.* [29] has suggested that when the Zeeman energy surpasses the thermal energy, there is field induced shifting of the two Weyl Fermi surfaces in each Dirac cone which is responsible for linear MR.

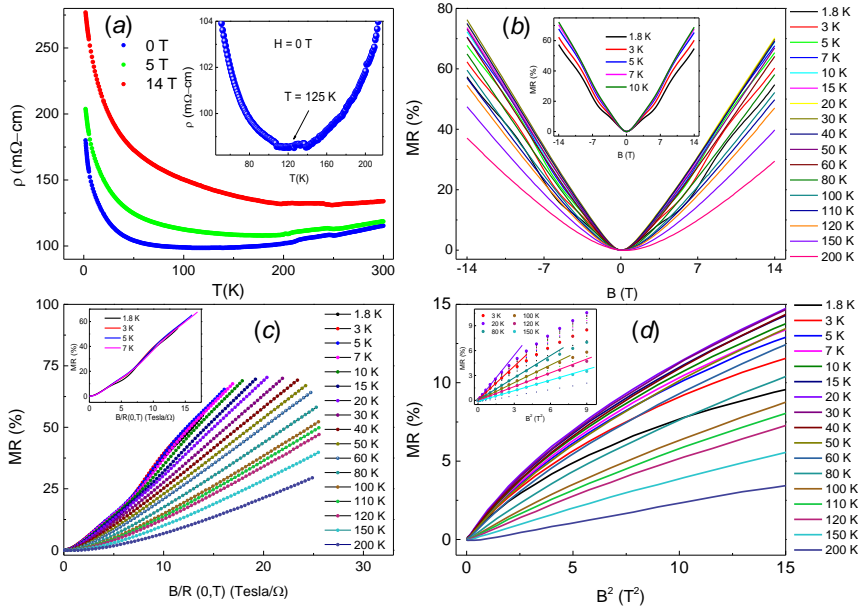

Figure 2. (a) The temperature dependence of $\rho(T)$ at *B = 0, 5, 14 Tesla*; Inset shows metal-semiconductor transition at *T ~ 125 K* for *B = 0 Tesla*. (b) The MR *(= (R(B) -R(0)) / R(0))* at different temperatures in the field range in *-14 to 14 Tesla* (inset shows quantum oscillations at *1.8, 3, 5, 7, 10 K*). (c) Kohler plots at different temperatures (inset shows merging of plots for *T ≤ 7 K*, on a single curve). (d) The variation of *ΔR/R* against $B^2$ where *ΔR/R ∝ $B^2$* criterion is obeyed for *B ≤2 Tesla*. Inset shows linear relation between MR and $B^2$ for few selected temperatures.

**THE UNUSUAL QUANTUM OSCILLATIONS**

Figure 3(a) shows the oscillations at 1.8, 3, 5, 7, and 10 K. In order to extract the oscillations from MR data, we subtracted the 4[th] order polynomials fit of the averaged MR (obtained from the MR values in positive and negative field directions) from the experimental data. As seen from the figure 3(a), distinct oscillations were observed for $T \leq 10\ K$ with clear peaks (maxima) at *3.2* and *9.0 Tesla*. Surprisingly these oscillations are periodic in *B* with the period $\Delta B$ of *2.92 Tesla*, but not in *1/B* (Figure 3 (a), (b)). The oscillations are usually associated with quantum interference effects and quantized Landau levels [18, 34]. The Landau level quantization give rise to Shubnikov de Haas (SdH) oscillations which are periodic in 1/B [18,34]. However, non- periodic oscillations in *1/B* similar to our compound have been observed in several compounds including some transition metal chalcogenide topological insulators [34-40]. It is worthy to mention that H. Wang *et al.* observed periodicity in *log B* instead of *1/B* or *B* for the *p*-type ZrTe$_5$ [18]. The log-periodic field dependence as a manifestation of DSI is suggested to originate from Effimov physics and intrinsic property of stoichiometric ZrTe$_5$ beyond the quantum limit [18]. The MR oscillations in Effimov physics appear due to change in densities of mobile carriers upon the formation and dissipation of Effimov bound states with the application of magnetic fields [18]. The periodicity of oscillations in B could be associated with size effects and geometrical confinements [35]. For the bulk materials, possibility of geometrical confinements is less likely, therefore fluctuation in carrier density or the change of Fermi surface cross section with the magnetic field could be the most likely reason for the non-periodictiy in *1/B* for our sample. The exchange of carriers between dominant plane bands (responsible for oscillations) and other bands comprising large reservoir of carriers could lead to the density fluctuation in the compound resulting in the unusual nature of oscillations in the compound. The Hall data supports our assumption of slight variation in carrier density with application of magnetic field (see top inset of figure S1 in Supplementary information).

We have shown $\Delta R$ vs. *1/B*, for T = 1.8 K in figure 3(b), which seems almost periodic in logarithmic scale. In order to compare our oscillation data with the literature and bring out the differences with the bipolar ZrTe$_5$, [23] we have analyzed our data by taking Fast Fourier Transform (FFT) of $\Delta R$ vs. *1/B*. The FFT performed on the oscillations gave single dominating peak at *F = 0.40 Tesla* along with other weak frequencies. The lower frequency in comparison to *F = 1.89 Tesla* for the reported single crystal [18] suggests the smaller size of Fermi surface and lower carrier concentration, which is consistent with the high value of Seebeck coefficient observed for the compound [2]. The amplitude of dominating frequency decreases with increasing temperature. The presence of additional weak peaks at higher frequencies along with the dominating peak at low frequency (inset of fig. 3(c)) shows that multiple bands contribute (weakly) to the transport properties of the compound. The similar type of behavior is observed in the single crystalline $ZrTe_5$ showing Fermionic Effimov states [18] but this behavior is in contrast to the observed in reference [23] for Te deficient bipolar $ZrTe_5$. The oscillation frequencies in our *p*-type $ZrTe_5$ can be predominantly divided into the three fundamental frequencies F1 (0.40 T), F2 (3.80T) and F3 (5.40T). F1 has multiple harmonics such as $4 \times F1$, $6 \times F1$, and $16 \times F1$. From the figure 3(c), it is evident that frequency F1 is dominant at low fields as well as high fields whereas frequency F2 and F3 do not have any harmonics or combination frequencies. As our preferred oriented $ZrTe_5$ shows *p*-type behavior in the whole *T* range, the frequency F1 should correspond to hole pockets. In the literature, it is observed that Fermi surface of bipolar $ZrTe_5$ comprises three pockets; one hole and two electron pockets and the volume of hole pocket is almost twice of the sum of the volumes of electrons pockets [23]. Our observation is consistent with presence of just three Fermi surfaces. It is interesting to note that in Te deficient bipolar $ZrTe_5$ all three frequencies have harmonics or inter-combination frequencies, but in our *p*-type $ZrTe_5$, combination frequencies are obtained for F1 only. To estimate the extremal Fermi surface cross-section area ($A_F$), we have used Onsager relation $F = \frac{\phi_0}{2\pi^2} A_F$ where $\phi_o$ is $2.067 \times 10^{-15}$ Wb [28]. The estimated Fermi surface cross sectional area $A_F$, Fermi wave vector $k_F$ ($A_F = \pi k_F^2$) and 2-D carrier density $n_{2D} \left(= \frac{k_F^2}{4\pi}\right)$ are given in table I.

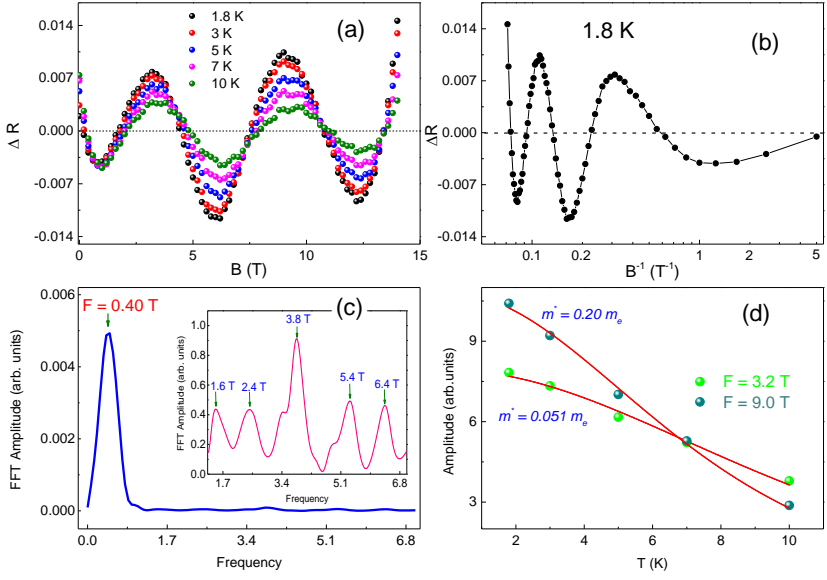

Figure 3. (a) SdH oscillations in MR for $T = 1.8, 3, 5, 7,$ and $10\ K$. (b) $\Delta R$ versus $1/B$ plot for $T = 1.8\ K$, (c) The frequency dependence of the FFT amplitude obtained from the FFT analysis of $\Delta R$ vs. $1/B$ curves. The inset shows the higher order harmonics in the zoomed scale. (d) The Lifshitz - Kosevich fit to the experimental data for the oscillation peak amplitude versus T for H = 3.20 and 9.0 Tesla.

| FFT Frequency F (Tesla) | Fermi cross-sectional area $A_F$ ($\times 10^{12}$ cm$^{-2}$) | Fermi wave vector ($k_F$) ($\times 10^5$ cm$^{-1}$) | 2 D carrier density $n_{2D}$ (cm$^{-2}$) |
|---|---|---|---|
| 0.4 | 0.382 | 3.49 | $9.68 \times 10^9$ |
| 1.6 | 1.53 | 6.98 | $3.88 \times 10^{10}$ |
| 2.4 | 2.29 | 8.54 | $5.72 \times 10^{11}$ |
| 3.8 | 3.63 | 10.75 | $9.08 \times 10^{11}$ |
| 5.4 | 5.16 | 12.82 | $1.29 \times 10^{12}$ |
| 6.4 | 6.11 | 13.94 | $1.53 \times 10^{12}$ |

The $A_F$ value at the dominating frequency (*0.40 T*) and other higher frequencies is one to two order lower in magnitude than for the bipolar ZrTe$_5$ crystal where $A_F$ ranges from *5.7 - 55.8 $\times 10^{12}$ cm$^{-2}$* for three different Fermi surfaces along three different axes [23]. The lower values of $A_F$ for *p*-type ZrTe$_5$ indicates that Fermi level is very close to the Dirac point. Similarly $k_F$ values observed for *p*-type ZrTe$_5$ are of same order as in bipolar ZrTe$_5$ crystal [23]. The areal carrier density $n_{2D}$ lies in the range *9.68 $\times 10^9$* to *1.53 $\times 10^{12}$ cm$^{-2}$* as shown in table and comparable to $2.87 \times 10^{11}$ cm$^{-2}$ of topological superconductor LuPdBi and $2.6 \times 10^{12}$ cm$^{-2}$ of thin flake ZrTe$_5$ [41-42]. In the fig 3(d), we have fitted oscillation amplitude vs. *T* for the peaks at *H = 3.2 T* and *9.0 T* using Lifshitz-Kosevich (L-K) expression [43] $\Delta R \propto \frac{2\pi^2 k_B T m^*/\hbar e B}{\sinh(2\pi^2 k_B T m^*/\hbar e B)}$, where ΔR is the oscillation amplitude. The (cyclotron) effective mass obtained from the fits are 0.051$m_e$ and 0.20 $m_e$ corresponding to the oscillation peaks at *H = 3.2* and *9.0 T* respectively. These values lies in the range of *0.028 $m_e$* to *0.845 $m_e$* as observed for the single crystal bipolar ZrTe$_5$ along different axis [23,44]. Considering the $k_F$ value at the dominating FFT frequency (*F = 0.4 T*), the calculated Fermi velocity ($v_F \sim \hbar k_F/m^*$) for effective mass *0.051 $m_e$* is *0.8 $\times 10^5$ m/s*, which is of the same order as for reported Te deficient single crystal [15] and stoichiometric (p-type) ZrTe$_5$ [18]. The estimated Fermi energy $E_F$ lies in the range *1.84 -7.26 meV* for $m^*$ = 0.051$m_e$ and 0.20$m_e$. Fitting for the Dingle temperature ($T_D$) with the formula $\Delta R \propto exp(2\pi^2 k_B T_D/\hbar \omega_c)$[35] where $\omega_c = eB/m^*$, and ln$\Delta R$ vs. *1/B* plots gives $T_D \sim$ *1.91 K*. This value is smaller than value reported from *4.47* to *11.5 K* for hole pockets but very close to $T_D \sim$ *1.83 -1.90 K* for the electron pockets along 'a' and 'c' axis [23]. The lifetime τ of carriers obtained from $T_D = \frac{\hbar}{2\pi k_B \tau} \approx 6.37 \times 10^{-13} s$, is of same order as for bipolar Te deficient ZrTe$_5$ single crystal [23] but one order higher than LuPdBi [41]. This gives the mean free path $l \sim v_F \tau$ as *50 nm* and the surface mobility $\mu_S \sim e\tau.m^*$ as *2.2 $\times 10^4$ cm$^2$V$^{-1}$s$^{-1}$*. The $\mu_S$ value is much higher than other TIs and Dirac semimetals such as Bi$_2$Te$_3$ (*10,200 cm$^2$V$^{-1}$s$^{-1}$* at *0.3 K*), LuPdBi (2100 cm$^2$V$^{-1}$s$^{-1}$ at 2.5 K), metallic bismuth tellurides sheets (6825±2100 cm$^2$V$^{-1}$s$^{-1}$), BiSbTe$_3$ (4490 cm$^2$V$^{-1}$s$^{-1}$), TlBiSSe (3500 cm$^2$V$^{-1}$s$^{-1}$) and Cd$_3$As$_2$ ($1.5 \times 10^4$ cm$^2$V$^{-1}$s$^{-1}$ at 300 K) [41, 45-49]. But our $\mu_S$ value is approximately half of the highest mobility of $\sim 4.4 \times 10^4$ cm$^2$V$^{-1}$s$^{-1}$ observed for the ZrTe$_5$ along 'a' axis and mean free path *l* is almost double of it (23 nm) [44]. Such a high value of surface mobility ensures the high quality of the sample and suggests that quantum oscillations originate from the surface states which are topologically protected, because in the bulk, mobilities (3000 and 1500 cm$^2$V$^{-1}$s$^{-1}$ along a- and c-axis) in the bulk ZrTe$_5$ [44] are very low in comparison to surface states. The surface contribution to overall conduction could be calculated by taking the ratio of surface state conductance $G_S$ to $R_{bulk}^{-1}$ (where $G_S = (e^2/h) \times k_F l$, and $R_{bulk}$ is the resistance at 1.8 K) [47]. The surface contribution is just 0.007 % of overall conduction which is again in agreement with quantum oscillations originating from the surface. The metallic parameter ($k_F l$) of our $ZrTe_5$ at dominating frequency comes around ~ 1.75 which is lower in comparison to $k_F l$ ~ 5 for metallic LuPdBi [41].

## SIGNATURE OF TOPOLOGICAL SURFACE STATES

Figure 4a shows the magnetization versus field curves in the field range *-7 to 7 Tesla* at different *T* from *1.8 to 380 K*. At low temperatures, the data show positive susceptibility (paramagnetic) at lower fields, and changes to negative (diamagnetic) at higher fields. On increase of *T*, the field range and magnitude of the positive susceptibility decreases. The positive susceptibility region decreases from *4 Tesla* at *1.8 K* to *130 Oe* at *380 K*. In the 3-D TIs, due to the spin-momentum locking the electron spin could align along the applied magnetic field direction due to singularity in spin orientation, which results in a low field paramagnetic peak in magnetic isotherm curves [12, 50]. Similar to *ZrTe₅* (*p*-type), *Sb₂Te₃*, *Bi₁.₅Sb₀.₅Te₁.₇Se₁.₃*, *Bi₂Se₃* and *Bi₂Te₃* [12,50], our compound also shows cusp like paramagnetic susceptibility (at low fields) which grows over diamagnetic floor (at higher fields) [12,50].

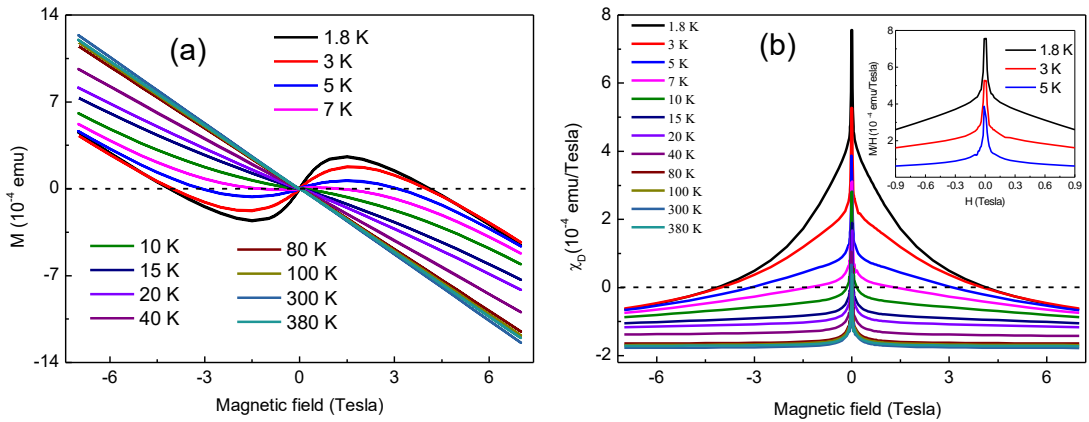

Figure 4: **(a)** Magnetization versus field data measured at *T = 1.8 to 380 K* in the field range *-7 to 7 Tesla*. **(b)** Field dependence of DC magnetic susceptibility ($\chi_D$) shows susceptibility cusp at low fields at several temperatures from *1.8 to 380 K*. Inset shows the zoomed curves for *T = 1.8, 3,* and *5 K*.

A. Pariari *et al.* have shown the cusp like paramagnetic susceptibility in *p*-type $ZrTe_5$ single crystal along *a*-axis in the low field range of $H \leq 0.2\ T$ from *2 to 350 K* and diamagnetic response at the higher fields [12]. The decrease of susceptibility (shown in fig 4b) with increasing *T* shows suppression of singular response of paramagnetic susceptibility due to thermal smearing. However in $ZrTe_5$, $Sb_2Te_3$, $Bi_2Se_3$ and $Bi_2Te_3$ single crystals the magnitude of paramagnetic peak is insensitive to the *T* [12,50]. The paramagnetic Dirac susceptibility ($\chi_D$), when chemical potential $\mu$ and *T* are set to zero, could be given as [50]

$$\chi_D(B) \cong \frac{\mu}{4\pi^2}\left[\frac{(g\mu_B)^2 \Lambda}{\hbar v_F} - \frac{2(g\mu_B)^3}{\hbar^2 v_F^2}|B| + \cdots\right]$$

where $g$ is the Lande factor, $v_F$ is the Fermi velocity and $\Lambda$ is the size of momentum space. The $\chi_D$ shows a cusp at low fields, and its value depends on the effective size of the momentum space $\Lambda$, and therefore controlled by the various factors such as hexagonal warping of the Dirac cone and by the bulk bands [12,50]. At low fields, $\chi_D$ is directly proportional to $\Lambda$, which is controlled by the details of bulk bands, which thus depend on $T$. The phenomenological description of low field paramagnetic response could be given in terms of effective Dirac bandwidth ($W = \hbar v_F \Lambda$) and field energy ($E_B = g\mu_B B$). The width of cusp is set by the condition $W \approx E_B$ [50]. The observed change in paramagnetic susceptibility with $T$ requires that the thermal energy $E_T$ ( $= k_B T$ ) should be higher than $W$ and $E_B$ which in turn depend on $\Lambda$ and $B$. As $v_F$, $B$ and $g$ ($2m_e/m_c \sim 40$ from the oscillation analysis) values are fixed, $\Lambda$ plays an important role for depression in amplitude of paramagnetic susceptibility with increasing $T$. It is worthy to mention that anisotropy also plays a great role in ZrTe$_5$ as single crystals are highly anisotropic in compare to preferred oriented ZrTe$_5$ [13]. The observed singularity in susceptibility at low fields is typical signature of 3-D TIs, arising from the sample surface due to the opening of a Zeeman gap at the Dirac point of the helical metal [12] and is universal to the 3-D TIs. The singularity is independent of the bulk carrier density, and predicts the existence of electronic states near the spin-degenerate Dirac point.

**CONCLUSIONS**

We have done detailed analysis of MR on the high quality preferred oriented $p$-type ZrTe$_5$ polycrystal which do not show the peak shape resistivity anomaly down to 1.8 K. We observed unusual quantum oscillations below 15 K which are periodic in $B$. The analysis of the MR electronic transport data and paramagnetic peak at lower fields in the Dirac susceptibility in magnetization data suggest the presence of highly conducting topologically protected surface states in the compound. The oscillations analysis and Kohler rule at low temperatures confirms the single band (hole) dominated transport properties in $p$-type ZrTe$_5$ in contrast to multiple band dominated bipolar ZrTe$_5$. The crossover from the quadratic (at low fields) to linear MR (at higher fields) behavior suggests that preferred oriented *p-type* ZrTe$_5$ behave like 3-D Weyl semimetal. Our results on the unusual quantum oscillation along with the reported log periodic oscillation suggests interesting and exotic hidden physics of the electronic transport in $p$-type ZrTe$_5$. Further studies are required to identify the exotic physics of quantum oscillations and topological features of $p$-type ZrTe$_5$.

**ACKNOWLEDGEMENT:** The authors acknowledge Advanced Material Research Center (AMRC), IIT Mandi for the experimental facilities. The financial support from the IIT Mandi and from the seed grant project IITMandi/SG/ASCY/29, and DST-SERB project YSS/2015/000814 is also acknowledged.

**REFERENCES**
[1] F. J. Disalvo, R. M. Fleming, and J. V. Waszczak, Phys. Rev. B **24**, 6 2935 (1981).
[2] M. K. Hooda, and C. S. Yadav, Appl. Phys. Lett. **111**, 053902 (2017).
[3] J. Niu *et al.*, Phys. Rev. B **95**, 035420 (2017).

[4] G. Manzoni, L. Gragnaniello, G. Autès, T. Kuhn, A. Sterzi, F. Cilento, M. Zacchigna, V. Enenkel, I. Vobornik, L. Barba *et al.*, Phys. Rev. Lett. **117**, 237601 (2016).

[5] G. Zheng, X. Zhu, J. Lu, W. Ning, H. Zhang, W. Gao, Y. Han, J. Yang, H. Du, K. Yang *et al.*, Phys. Rev. B **96**, 121401(R) (2016).

[6] Xiang-Bing Li *et al.*, Phys. Rev. Lett. **116**, 176803 (2016).

[7] Y. Liu, X. Yuan, C. Zhang, Z. Jin, A. Narayan, C. Luo, Z. Chen, L. Yang, J. Zou, X. Wu *et al.*, Nat. Commun. **7**, 12516 (2016).

[8] Z.-G. Chen, R. Y. Chen, R. D. Zhong, J. Schneeloch, C. Zhang, Y. Huang, F. Qu, R. Yu, Q. Li, G. D. Gu, and N. L. Wang, Proc. Natl. Acad. Sci.**114**, 816 (2017).

[9 L. Moreschini, J. C. Johannsen, H. Berger, J. Denlinger, C. Jozwiak, E. Rotenberg, K. S. Kim, A. Bostwick, and M. Grioni, Phys. Rev. B **94**, 081101(R) (2016).

[10] H. Chi, C. Zhang, G. Gu, D. E. Kharzeev, Xi Dai, and Q. Li, New J. Phys. **19**, 015005 (2017).

[11] H. Xiong, J. A. Sobota , S.-L. Yang, H. Soifer, A. Gauthier, M.-H. Lu, Y.-Y. Lv, S.-H. Yao, D. Lu, M. Hashimoto *et al.*, Phys. Rev. B **95**, 195119 (2017).

[12] A. Pariari, and P. Mandal, Scientific Reports **7**, 40327 (2017).

[13] P. Shahi, D. J. Singh, J. P. Sun, L. X. Zhao, G. F. Chen, J. Q. Yan, D. G. Mandrus, J. G. Cheng, arxiv:1611.06370v1 (2016).

[14] R. Wu *et al.*, Phys. Rev. X **6**, 021017 (2016).

[15] G. Zheng *et al.*, Phys. Rev. B **93**, 115414 (2016).

[16] Z. Fan, Qi-Feng Liang, Y. B. Chen, Shu-Hua Yao, and J. Zhou, Scientific Reports **7**, 45667 (2017).

[17] Q. Li *et al.*, Nat. Phys. **12**, 550 (2016).

[18] H. Wang *et al.*, arxiv:1704.00995 (2017).

[19] C. Zhang *et al.*, Phys. Rev. B **94**, 041203(R) (2015).

[20] N. Ramakrishnan, M. Milletari, and S. Adam, Phys. Rev. B **92**, 245120 (2015).

[21] K. Ziegler, Eur. Phys. J. B. **89**, 268 (2016).

[22] J. M. Ziman, *Electrons and Phonons*, (Oxford, UK, Clarendon Press, 2001).

[23] G. N. Kamm *et al.*, Phys. Rev. B **31**, 7617 (1985).

[24] C. Shekhar *et al.*, Nat. Phys. **11**, 645 (2015).

[25] T. Liang *et al.*, Nat Mater. **14**(3), 280 (2015).

[26] R. Xu *et al.*, Nature **390**, 57 (1997).

[27] W. Zhang *et al.*, Phys. Rev. Lett. **106**, 156808 (2011).

[28] L. P. He *et al.*, Phys. Rev. Lett. **113**, 246402 (2014).

[29] J. Feng *et al.*, Phys. Rev. B **92**, 081306(R) (2015).

[30] C.-L. Zhang *et al.*, Phys. Rev. B **95**, 085202 (2017).

[31] P. N. Argyres, J. Phys. Chem. Solids **4**, 19 (1958).

[32] A. Abrikosov, Phys. Rev. B **58**, 2788 (1998).

[33] A. A. Abrikosov, Sov. Phys. JETP **29**, 746 (1969).

[34] Yu Pan, Ph.D. thesis, Amsterdam University, 2016 (Chapter 5, page 61-63).

[35] V. Dhurairaj *et al.*, Phys. Rev. B **73**, 054434 (2006).

[36] C. Martin *et al.*, Phys. Rev. B **90**, 201204(R) (2014).

[37]. W. Shim, J. Ham, J. Kim, and W. Lee, Appl. Phys. Lett. **95**, 232107 (2009).

[38] F.-X. Xiang, X.-L. Wang, M. Veldhorst, S.-X. Dou, and M. S. Fuhrer, Phys. Rev. B **92**, 035123 (2015).

[39] S. Harashima, C. Bell, M. Kim, T. Yajima, Y. Hikita, and H. Y. Hwang, Phys. Rev. B **88**, 085102 (2013).

[40] F. Y. Yang, K. Liu, K. Hong, D. H. Reich, P. C. Searson, C. L. Chien, Y. L.-Wang, K. Yu-Zhang, and Ke Han, Phys. Rev. B **61**, 6631 (2000).

[41] O. Pavlosiuk, D. Kaczorowski, and P. Wisniewski, Scientific Reports **5**, 09158 (2015).

[42] W. Yu *et al.*, Scientific Reports **6**, 35357 (2016).

[43] K. S. Novoselov, A. K. Geim, S. V. Morozov, D. Jiang, M. I. Katsnelson, I. V. Grigorieva, S. V. Dubonos, and A. A. Firsov, Nature **438**, 197 (2005).

[44] G. Qiu *et al.*, Nano Lett. **16**, 7364−7369 (2016).

[45] M. Novak *et al.*, Phys. Rev. B **91**, 041203(R) (2015).

[46] D.-X. Qu, Y. S. Hor, J. Xiong, R. J. Cava, and N. P. Ong, Science **329**, 821 (2010).

[47] T. Chen *et al.*, Eur. Phys. J. D **67**, 75 (2013).

[48] F.-X. Xiang, X.-L. Wang, and S.-X. Dou, arXiv:1404.7572 (2014).

[49] Z. Wang, H. Weng, Q. Wu, Xi Dai, and Z. Fang, Phys. Rev. B **88**, 125427 (2017).

[50] L. Zhao *et al.*, Nat. Mater. **13**, 580 (2014).

# SUPPLEMENTARY INFORMATION

# Unusual magnetoresistance oscillations in preferentially oriented *p*-type polycrystalline ZrTe$_5$

M. K. Hooda and C. S. Yadav

School of Basic Sciences, Indian Institute of Technology Mandi, Mandi-175005 (H.P.) India

Figure S1 shows the Hall resistivity ($\rho_{xy}$) measured over the temperature (*T*) range 3 to 300 K in the field range 0 to 14 *Tesla*. The $\rho_{xy}$ shows the one type of carrier (hole) dominance in the whole *T* range. The trend observed in $\rho_{xy}$ of our preferentially oriented polycrystal is consistent with the single crystal Hall data reported by A. Pariari *et al.* in the field range 0 to 9 *Tesla* [1]. The hole dominance in whole *T* range ensure *p*-type nature of the compound and also the absence of resistivity anomaly which arises from the switch over from *p*-type to *n*-type carriers. The $\rho_{xy}$ shows the signature (very weak) of oscillations from 3 to 10 K like the magnetoresistance data. Above 10 K with disappearance of oscillations, the $\rho_{xy}$ follows the convex slope up to 150 K and at 300 K, the curve becomes linear. We have calculated the carrier concentration (*n*) using single band model fitting the linear region at lower fields (0 to 4 *Tesla*) and higher fields (7 to 14 *Tesla*) below 300 K as shown in the top inset of figure S1. The $\rho_{xy}$ curve behavior in the intermediate field range (*4 < B < 7 Tesla*) is quite complicated and does not follow linear behavior. The calculated *n* varies from 0.92 × 10$^{16}$ to 1.92× 10$^{17}$ *cm*$^{-3}$ for *3 K ≤T  ≤300 K* which is in agreement with previous reports about *p*-type ZrTe$_5$ crystal [1, 2]. There is variation in *n* with change of *T* and this confirms possibility of slight variation in carrier density which disturbs the periodicity in 1/B in the oscillations. The variation in *n* with *T* is similar to the reported by H. Wang et al. [2].

**References**
[1] A. Pariari, and P. Mandal, Scientific Reports **7**, 40327 (2017)
[2] H. Wang *et al.*, arxiv:1704.00995 (2017).

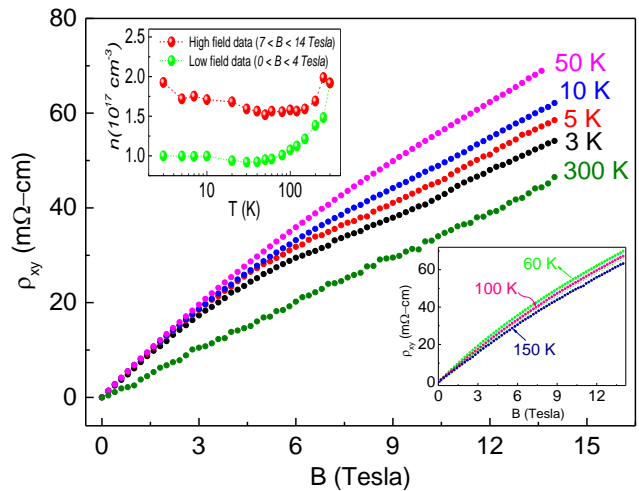

Figure S1. Hall resistivity ($\rho_{xy}$) from 3 K to 300 K in the field range of 0 to 14 *Tesla*. Top inset shows the variation of carrier density n with *T* from 3 K to 300 K. Bottom inset shows $\rho_{xy}$ at 60 K, 100 K and 150 K.

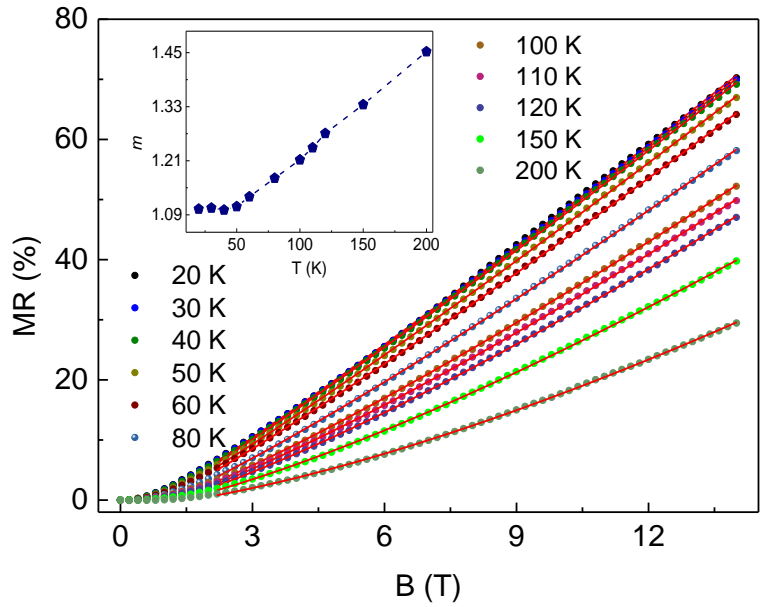

Figure S2. MR curve fit with the equation $y = y_0 + aB^m$ for $B > 2$ *Tesla*. The value of the exponent '*m*' vary from *1.1* at *20 K* to *1.45* at *200 K*. The red line shows fitting to MR data above 2 Tesla. The quadratic field dependence of MR is observed up to 2 *Tesla*. Top inset shows the variation of exponent *m* with temperature from 20 K to 200 K.

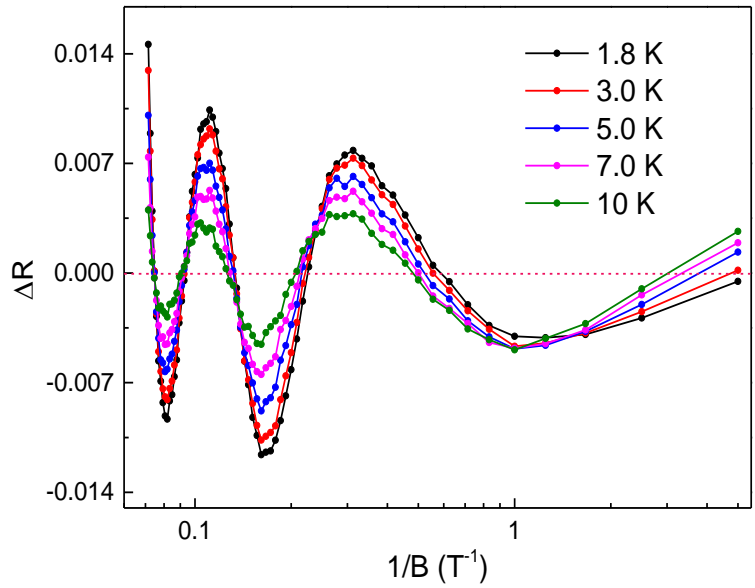

Figure S3: MR oscillations against 1/B for temperature range 1.8 to 10 K.